\documentclass[twocolumn,aps,showpacs,prb,tightenlines,amsmath,amssymb,superscriptaddress]{revtex4}
\usepackage{graphicx}
\usepackage{amssymb}
\usepackage{dcolumn}
\usepackage{amsmath}
\usepackage{bm}

\usepackage{colordvi}

\begin{document}

\title{A peak in density dependence of electron spin 
relaxation time in $n$-type bulk GaAs in metallic regime}

\author{K.\ Shen}
\affiliation{Hefei National Laboratory for Physical Sciences at
Microscale,
University of Science and Technology of China, Hefei,
Anhui, 230026, China}
\affiliation{Department of Physics,
University of Science and Technology of China, Hefei,
Anhui, 230026, China}
\altaffiliation{Mailing address}

\date{\today}

\begin{abstract}
We demonstrate that the peak in the density dependence of 
electron spin relaxation time in $n$-type bulk GaAs in the metallic regime
predicted by Jiang and Wu [Phys. Rev. B {\bf 79}, 125206 (2009)] has been
realized experimentally in the latest work by Krau\ss\ {\em et al}. 
[arXiv:0902.0270].
\end{abstract}

\pacs{72.25.Rb, 72.25.Fe, 71.70.Ej, 78.20.Nv}

\maketitle

Carrier spin lifetime in semiconductors has attracted much attention
because of its essential role for the application of spintronic
devices.\cite{spintronic} As reported, the electron spin lifetime can
be strikingly long, even exceed 
300~ns,\cite{dzhioev} in $n$-type bulk GaAs at liquid helium
temperature.\cite{dzhioev2,dzhioev3,kikkawa,furis} 
Due to the localized 
effect at such a low temperature, a 
very unusual behavior arises from the donor concentration dependence
of the spin lifetime, where two maxima appear.\cite{dzhioev3} The
one at relatively low doping concentration comes from 
the interplay of the hyperfine 
interaction and the anisotropic exchange interaction of donor-bound
electrons, while the other one originates 
from the metal-to-insulator transition (MIT).\cite{shklovskii} In the
metallic regime, the D'yakonov-Perel'(DP) mechanism associated with the spin
procession under the momentum-dependent effective magnetic
field\cite{dp} together with spin-conserving scattering is
recognized as the dominant electron spin relaxation mechanism in
$n$-type semiconductors.\cite{dzhioev3,jiang,jiang2,wu2,zhou1} Very
recently, Jiang and Wu studied the electron spin dynamics in bulk GaAs in
{\em metallic} regime from a fully microscopic 
kinetic spin Bloch equation (KSBE)
approach\cite{wu1,wu2,zhou2} and predicted a non-monotonic donor
concentration dependence of the spin relaxation time, where 
the maximum spin relaxation time occurs at the crossover of the
degenerate regime to the non-degenerate one, i.e.,
the corresponding Fermi temperature $T_F$ at the peak comparable
to the temperature of the electron plasma.\cite{jiang} 
In a recent experiment, a peak in doping density dependence of electron
spin relaxation time was also reported in optically excited $n$-type
bulk GaAs at 4~K,\cite{krauss} where the spin relaxation times are only several hundred
picoseconds, shorter than the typical spin lifetime of the localized electron
by two orders of magnitude.\cite{dzhioev}
This reveals that the peak should be associated with the spin 
dynamics in the {\em metallic}
regime instead of the localized effect.\cite{krauss}
In that paper, by simply solving the KSBEs, the maximum spin relaxation time 
was interpreted as the interplay of the elastic impurity-carrier
scattering, the inelastic carrier-carrier/phonon scattering, and the
influence of screening.

In the present letter, we show that the physics of the
peak is nothing but the one predicted by Jiang and Wu in Ref.\
\onlinecite{jiang}. One would notice that the 
corresponding Fermi temperature at the experimental peak is around 90~K, which
is much higher than the lattice temperature.  
Even though the experiment was performed at 4~K,
we will show that due to the laser-induced hot-electron effect, the electron
temperature can be higher than the lattice temperature, and the
underlying physics of the peak
is coincident with that proposed by Jiang and Wu, i.e., the
crossover of the degenerate regime to the non-degenerate one.

We study electron spin dynamics in $n$-type bulk GaAs in metallic
regime from the fully microscopic KSBEs derived via the nonequilibrium Green
function method,\cite{wu1,wu2,bronold,haug}
\begin{equation}
\partial_t\rho_{\bf k}=\partial_t\rho_{\bf k}|_{\rm
  coh}+\partial_t\rho_{\bf k}|_{\rm scat},
\label{eq1}
\end{equation}
where $\rho_{\bf k}$ is the density matrix of the electron with
momentum ${\bf k}$. $\partial_t\rho_{\bf k}|_{\rm coh}$ is the
coherent term, which describes the spin precession under 
the effective magnetic field originating from the Dresselhaus spin-orbit coupling
(SOC)\cite{dresshaus} and Coulomb Hartree-Fock 
(HF) terms.\cite{wu2,jiang} 
$\partial_t\rho_{\bf k}|_{\rm scat}$ represents the 
scattering term, including all the relevant scatterings, such as
the electron-impurity, electron-phonon, electron-electron, and
electron-hole scatterings. Both lattice-temperature longitudinal
optical (LO) phonon
and acoustic (ac) phonon are included in our KSBEs. The details of
Eq. (\ref{eq1}) can be found in Ref. \onlinecite{jiang}. 

\begin{figure}
\centering
\includegraphics[width=7.5cm]{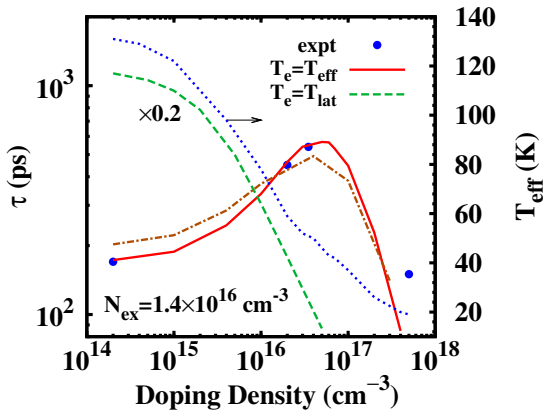}
\caption{ (Color online)  Electron spin 
  relaxation times from the calculation via the KSBE approach (red
  solid curve) and from the experiment\cite{krauss} (blue $\bullet$) as 
  function of the doping density. The green dashed curve shows the
  results without hot-electron effect. In the
  calculation, the Dresselhaus
SOC $\gamma_D=26$~eV$\cdot$\AA$^{3}$,\cite{jancu}
  photo-excited carrier density $N_{ex}=1.4\times10^{16}$~cm$^{-3}$, 
and the photon energy $\hbar\nu=1.55$~eV. The
  hot-electron temperature used in the computation is plotted as the blue
  dotted curve (Note the scale is on the right hand side of the
frame). The chain curve is the calculated spin relaxation time with a fixed
hot-electron  temperature  80~K, and $N_{ex}=6\times10^{15}$~cm$^{-3}$.}
\label{fig1}
\end{figure}

For optically excited setup in $n$-type GaAs, the total electron density
$N_e=N_{ex}+N_D$, with $N_{ex}$ and $N_D$ representing those from
 photo-excited process and
$n$-type dopant, respectively. $N_D$ equals to the impurity density $N_i$ for
Si-doped GaAs.\cite{kikkawa} Initially, $N_D$ is in the thermal
equilibrium state at lattice temperature $T_L$, and the photo-excited
carriers are in Gaussian distribution with the center energy
$\hbar\nu=1.55$~eV in this work.
The optimal polarization of $N_{ex}$ generated by a circularly
polarized pump pulse is 50~\%, according to the optical
selection rule.
Experimental and theoretical studies on ultrafast hole spin dynamics
reveal that holes lose their polarization information within
sub-picosecond in bulk GaAs.\cite{hilton,krauss2}
Therefore, it is reasonable to employ a time-independent thermal equilibrium hole
distribution with the density $N_h=N_{ex}$ in the time scale of interest 
here (several hundred picoseconds). We take the hole temperature equals
to the hot-electron temperature approximately. The
time evolution of the electron polarization signals $\langle
s_z\rangle=\sum_{\bf k}{\rm Tr}(\rho_{\bf k} {\bf s}_z)$ is obtained
by numerically solving KSBEs.

In Fig.\ \ref{fig1}, the electron spin relaxation time
is plotted as a function of the doping density
with the photo-excited electron density $N_{ex}=1.4\times10^{16}$ ${\rm
  cm}^{-3}$ as the red solid curve. It is seen that the spin
relaxation time presents a non-monotonic doping-density dependence
with the maximum achieved at around $N_{D}=5\times10^{16}$ ${\rm
  cm}^{-3}$, which is in good agreement with the experimental data (blue
$\bullet$).\cite{krauss}
This feature can be
explained by $\tau\sim 1/[\langle |{\bf
  \Omega}({\bf k})|^2-\Omega_z^2({\bf k})\rangle
\tau_p^\ast]$ in the strong scattering
limit,\cite{spintronic,jiang,wu2,glazov, leyland, harley} 
in which $\bf \Omega$ represents the
inhomogeneous effective
magnetic field and $\tau_p^\ast$ stands for
 the averaged effective momentum scattering
time. From our calculation, we find that the momentum relaxation process
of photo-excited carriers is completed  within a few picoseconds, 
leaving the electron distribution as a Fermi distribution
with the hot-electron temperature different from the lattice
one. We plot the hot-electron temperature $T_{\rm
  eff}$ as a function of the doping 
density as the blue dotted curve. This hot-electron temperature is exactly the
one used in the computation of spin relaxation time. 
We find that the hot-electron
temperature is beyond 100~K at low doping density, much higher than
 the lattice temperature
$T_{\rm lat}$, and decreases significantly with  the increase of  doping
density. Therefore, the system is non-degenerate (degenerate) in
the low (high) doping regime.
According to Ref.\ \onlinecite{jiang}, in the non-degenerate regime, i.e.,
$T_{\rm e}\gg T_F$, the inhomogeneous broadening $\langle|{\bf 
  \Omega}({\bf k})|^2-\Omega_z^2({\bf k})\rangle$ is insensitive to
the doping density, and the electron-electron scattering time
$\tau_p^{ee}$, electron-hole scattering time $\tau_p^{eh}$, and the
electron-impurity scattering time $\tau_p^{ei}$ all decrease
with increasing doping density. As a result, the spin
relaxation time $\tau$ increases with the doping density according to the
motional narrowing relation. 
Contrarily, the increase of the inhomogeneous
broadening becomes more efficient than the change of
$\tau_p^\ast$, and hence reduces $\tau$ in the degenerate
regime.\cite{jiang} The crossover  between 
the degenerate  and non-degenerate regimes can be estimated by the
Fermi temperature $T_F$.\cite{jiang} 
At the peak ($N_{D}\sim 5\times10^{16}$ ${\rm cm}^{-3}$), $T_F$ ($\sim 90$~K)
is comparable to the hot-electron
temperature ($\sim 45$~K), which is coincident with the
theory of Ref.\ \onlinecite{jiang}. 
For comparison, we also plot the results without the hot-electron effect,
i.e., $T_e=T_{\rm lat}$, as the
green dashed curve. It is seen that the spin relaxation time is much longer than
the result with the hot-electron effect at low
doping density and decreases monotonically with increasing the doping
density. The underlying physics of the short spin relaxation time due to the
hot-electron effect in the low doping regime is that the high hot-electron
temperature  drives the electrons to the
states with a larger momentum and hence enhances the inhomogeneous
broadening. Therefore the spin relaxation time decreases drastically.
The monotonic decrease of the spin relaxation time is because that
the electron-hole plasma is strongly degenerate in the entire range of doping density due to
the low temperature. 

One notices that the hot-electron temperature at the peak is
about one half of $T_F$, instead of $T_e\approx T_F$ as predicted in Ref.\
\onlinecite{jiang}. This results from
the varied hot-electron temperature at different doping densities. In
the low doing regime, the high hot-electron temperature reduces the spin
relaxation time, which leads the peak to the relatively high
doping regime. The brown chain curve shows the calculation with a fixed 
hot-electron temperature at 80~K, where the relation $T_e\approx T_F$ at the peak
is  satisfied.\cite{jiang}

To examine the role of the external magnetic field, we turn on the
magnetic field in the
coherent term in Eq.\ (\ref{eq1}).
We choose $B=4$~T in the Voigt configuration following the experimental
scheme.\cite{krauss} We find that the electron spin relaxation time determined
by the DP mechanism is only marginally different (not shown) from 
the zero-field results, which agrees with the previous
experiment.\cite{kikkawa}

In summary, we have investigated the electron spin dynamics from the fully
microscopic KSBEs in $n$-type bulk GaAs.
 We demonstrate that the peak observed in the latest
experiment\cite{krauss} gives  the strong  evidence of the prediction
by Jiang and Wu\cite{jiang}  via
analyzing the laser induced hot-electron effect.

The author would  like to thank M. W. Wu
for proposing the topic as well as directions during the whole investigation.
He would also like to thank J. H. Jiang and H. C. Schneider for
helpful discussions.
This work was supported by the Natural Science Foundation of China
under Grant No.\ 10725417, the National Basic Research Program of China
under Grant No.\ 2006CB922005 and the Knowledge Innovation Project of
Chinese Academy of Sciences.

\end{document}